\documentclass[preprint,showpacs,preprintnumbers,amsmath,amssymb]{revtex4}
\usepackage{graphicx}
\usepackage{dcolumn}
\usepackage{bm}

\begin{document}


\title{Clustering features of $^9$Be, $^{14}$N, $^7$Be, and $^8$B nuclei in relativistic
 fragmentation}

\author{D.~A.~Artemenkov}
  \email{artemenkov@lhe.jinr.ru}  
   \affiliation{Joint Insitute for Nuclear Research, Dubna, Russia}

\author{T.~V.~Shchedrina}
   \affiliation{Joint Insitute for Nuclear Research, Dubna, Russia}

\author{R.~Stanoeva}
   \affiliation{Joint Insitute for Nuclear Research, Dubna, Russia} 
 
 \author{P.~I.~Zarubin}
     \email{zarubin@lhe.jinr.ru}    
    \homepage{http://becquerel.lhe.jinr.ru}
   \affiliation{Joint Insitute for Nuclear Research, Dubna, Russia}

\date{\today}

\begin{abstract}
Recent studies of clustering in light nuclei with an initial
 energy above 1 A~GeV in nuclear treack emulsion are overviewed.
 The results of investigations
 of the relativistic $^9$Be nuclei fragmentation in emulsion, which entails the production
 of He fragments, are presented. It is shown that most precise angular measurements provided
 by this technique play a crucial role in the
 restoration of the excitation spectrum of the $\alpha$ particle sysytem.
 In peripheral interactions $^9$Be nuclei are dissociated practically totally 
through the 0$^+$ and 2$^+$ states
 of the $^8$Be nucleus.\par
\indent The results of investigations of the dissociation of a $^{14}$N nucleus
 of momentum 2.86~A~GeV/c in emulsion are presented as example of more complicated system. 
The momentum and correlation
 characteristics of $\alpha$ particles for the $^{14}$N$\rightarrow$3$\alpha+X$
 channel in the laboratory system and the rest systems of 3$\alpha$ particles were
 considered in detail.\par 
 \indent Topology of charged fragments produced in peripheral relativistic dissociation of 
radioactive $^8$B, $^7$Be nuclei
 in emulsion is studied.\par
\end{abstract}

\maketitle
\section{\label{sec:level1}Introduction}
\indent  The peripheral fragmentation of light relativistic nuclei can serve as a 
source of information about
 their exitations above particle decay thresholds including many-body final states. 
The interactions of this type are provoked either in electromagnetic
 and diffraction processes, or in nucleon collisions at small overlapping of the
 colliding nucleus densities. A fragmenting nucleus gains an excitation spectrum
 near the cluster dissociation thresholds. In the kinetic region of fragmentation
 of a relativistic nucleus there are produced nuclear fragment systems the total
 charge of it is close to the parent-nucleus charge. A relative intensity of
 formation of fragments of various configurations makes it possible to estimate the
 importance of different cluster modes.\par
 \indent The opening angle of the relativistic fragmentation cone is determined by
 the Fermi-momenta of the nucleon clusters in a nucleus. Being normalized to the
 mass numbers they are concentrated with a few percent dispersion near the normalized
 momentum of the primary nucleus. When selecting events with dissociation of a
 projectile into a narrow fragmentation cone we see that target-nucleus non-relativistic
 fragments either are absent (\lq\lq white\rq\rq stars in Ref.\cite{Andreeva05}),
 or their number is insignificant. The target fragments are easily separated from
 the fragments of a relativistic projectile since their fraction in the angular
 relativistic fragmentation cone is small and they possess non-relativistic
 momentum values.\par
 \indent In the peripheral fragmentation of a relativistic nucleus with charge Z 
 the ionization induced by the fragments can decrease down to a factor Z, while the
 ionization per one track -- down to Z$^2$. Therefore experiment should provide
 an adequate detection range. In order to reconstruct an event, a complete kinematic
 information about the particles in the relativistic fragmentation cone is needed
 which, e.g., allows one to calculate the invariant mass of the system.
 The accuracy of its estimation decisively depends on the exactness of the track
 angular resolution. To ensure the best angular resolution, it is necessary that
 the detection of relativistic fragments should be performed with the best spacial
 resolution.\par
\begin{figure}
    \includegraphics[width=5in]{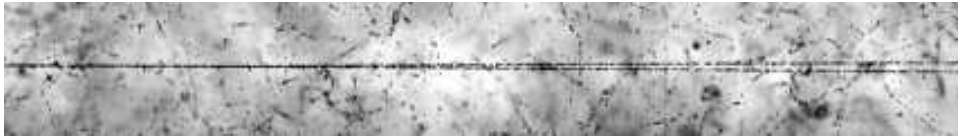}
    \caption{\label{fig:1} An event of the type of \lq\lq white\rq\rq star from the
 fragmentation of a relativistic $^9$Be nucleus into two He
 fragments in emulsion. The photograph was obtained on the
 PAVIKOM(FIAN)~complex.}
    \end{figure}
 \indent The nuclear emulsion technique, which underlies the BECQUEREL project at the
 JINR Nuclotron \cite{Web}, well satisfies the above-mentioned requirements.
 It is aimed at a systematic search for peripheral fragmentation modes with
 statistical provision at a level of dozens events, their classification and
 angular metrology. Emulsions provide the best spacial resolution (about 0.5~$\mu$m)
 which allows one to separate the charged particle tracks in the three-dimensional
 image of an event within one-layer thickness (600~$\mu$m) and ensure a high
 accuracy of angle measurements. The tracks of relativistic H and He nuclei are
 separated by sight. As a rule, in the peripheral fragmentation of a light nucleus
 its charge can be determined by the sum of the charges of relativistic fragments.
 Multiple-particle scattering measurements on the light fragment tracks enable one
 to separate the H and He isotopes. The analysis of the products of the relativistic
 fragmentation of neutron-deficient isotopes has some additional advantages owing
 to a larger fraction of observable nucleons and minimal Coulomb distortions.
 Irradiation details and a special analysis of interactions in the BR-2 emulsion
 are presented in Ref. \cite{Adamovich99, Adamovich04}. In what follows, we give
 the first results of the study of the $^9$Be,$^8$B, $^7$Be $^{14}$N nuclei fragmentation
 with a few A~GeV energy which are obtained with the use of a part of the material
 analyzed.\par
\section{\label{sec:level2}Fragmentation of $^9$Be nuclei}
\indent The $^9$Be nucleus is a loosely bound n+$\alpha$+$\alpha$ system.
 The energy threshold of the $^9$Be$\rightarrow$n+$\alpha$+$\alpha$ dissociation
 channel is 1.57~MeV. The study of the $^9$Be fragmentation at relativistic
 energies gives the possibility of observing the reaction fragments, which are the
 decay products of unbound $^8$Be and $^5$He nuclei.\par
 \indent The method of nuclear emulsions used in the present paper allows one to
 observe the charged component of the relativistic $^9$Be$\rightarrow$2He+n
 fragmentation channel. Owing to a good angular resolution of this method it is
 possible to separate the $^9$Be fragmentation events, which accompanied by
 the production of an unstable $^8$Be nucleus with its subsequent breakup to two a
 particles. In this case, the absence of a combinatorial background (of three and
 more $\alpha$ particles) for $^9$Be, which is typical for heavier N$\alpha$ nuclei $^{12}$C
 and $^{16}$O makes it possible to observe distinctly this picture.\par
\indent Nuclear emulsions were exposed to relativistic $^9$Be nuclei at the JINR
 Nuclotron. A beam of relativistic $^9$Be nuclei was obtained in the
 $^{10}$B$\rightarrow^9$Be fragmentation reaction using a polyethylene target.
 The $^9$Be nuclei constituted about 80\% of the beam, the remaining 20\% fell on
 Li and He nuclei.\cite{arXiv9Be}\par
 \indent Events were sought by microscope scanning over the emulsion plates.
 In total 362 events of the $^9$Be fragmentation involving
 the two He fragment production in the forward fragmentation cone within a polar
 angle of 6$^{\circ}$(0.1~rad) were found. 
 The requirement of conservation of the fragment
 charge in the fragmentation cone was fulfilled for the detected events. In event selection 5 - 7 tracks of various types 
were allowed
 in a wide (larger than 6$^{\circ}$) cone to increase statistics. An example of the
 $^9$Be$\rightarrow$2He fragmentation event in emulsion is given
 in Fig.~\ref{fig:1} \cite{Web}. This event belongs to the class of \lq\lq white\rq\rq stars
 as far as it contains neither target nucleus fragments, nor produced mesons. This event sample includes 144 \lq\lq white \rq\rq stars.
 The angles of the tracks in emulsion for the detected events were obtained using a fine measuring
 microscope. Angular measurements for the 362 events were carried out with an
 accuracy not worse then 4.5$\times$10$^{-3}$~rad.\par
 \indent In analyzing the data both He fragments observed in the
 $^9$Be$\rightarrow$2He+n channel were supposed to be a  particles. This assumption
 is motivated by the fact that at small angles the $^9$Be$\rightarrow2^4$He+n fragmentation channel with an energy threshold of
 1.57~MeV must dominate the $^9$Be$\rightarrow^3$He+$^4$He+n channel whose energy
 threshold is 22.15~MeV. The $^3$He fraction will not exceed a few percent in this
 energy range \cite{Belaga96} and all the He fragments in the detected events may be thought of
 as $\alpha$ particles.\par
\begin{figure}
    \includegraphics[width=5in]{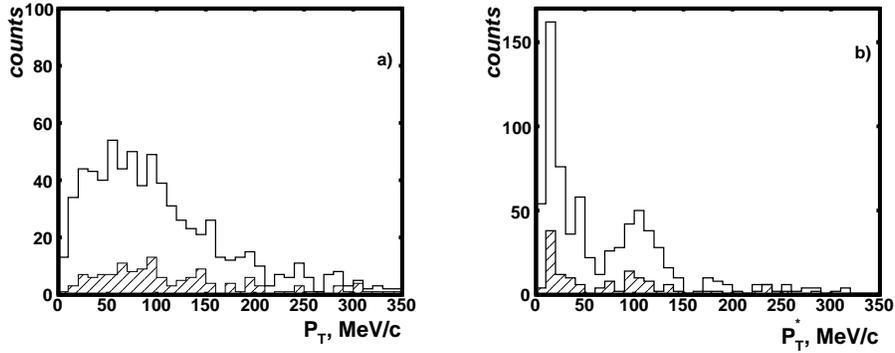}
    \caption{\label{fig:2} The P$_{T}$ transverse momentum distribution  of $\alpha$ particles
 in the laboratory system (a), and the P$^*_{T}$ momentum distribution in the c.m.s. of an
 $\alpha$ particle pair (b). The outer contour corresponds to all events. The inner histogram is obtained
 for events, which accompanied by protons recoil of emulsion target (dashed area).}
 \end{figure}
\indent In Fig.~\ref{fig:2}a  the P$_{T}$ transverse momentum distribution  of $\alpha$
 particles in the laboratory frame of reference is calculated without the account of
 particle energy losses in emulsion  by the equation  
\begin{equation}
{P_{T}=p_{0}\cdot A\cdot sin\theta }\label{eq1}
\end{equation}
where p$_{0}$, A and $\theta$ are the momentum per nucleon, the fragment mass
 and  the polar emission angle, respectively. The outer contour corresponds to all events.
The inner histogram is obtained for events accompanied by protons recoil of emulsion target (dashed area).
 The mean value of the transverse momentum for the total event sample
 in the laboratory system is equal to $<P_{T}>\approx$103 MeV/c with FWHM $\sigma\approx$72 MeV/c. 
 This may be an indication of the fact that the experimental data are not
 of the same kind which can be pronounced when going over to the c.m.s. of two $\alpha$ particles.\par
\indent  The P$^*_{T}$ transverse momentum distribution of $\alpha$ particles in the c.m.s.
 of two $\alpha$ particles described by the equation
\begin{equation}
   {\bf P}^{*}_{Ti}\cong {\bf P}_{Ti}- \frac{\sum_{i=1}^{n}{\bf P}_{Ti}}{n_{\alpha}}\label{eq2} 
 \end{equation}
where P$_{Ti}$ is the transverse momentum  of an i-th $\alpha$ particle in the laboratory
 system n$_{\alpha}$=2  is given in Fig.~\ref{fig:2}b.  There is observed a grouping of
 events around  two peaks with the values $<P^*_{Ti}>\approx$24~MeV/c and $<P^*_{Ti}>\approx$101~MeV/c.
 In Ref~\cite{Avetyan96} the appropriate mean values of the $\alpha$ fragment transverse momenta
 are  $<P^*_{Ti}>\approx$121~MeV/c for $^{16}$O$\rightarrow$4$\alpha$,$<P^*_{Ti}>\approx$141~MeV/c
 \cite{Belaga95} for $^{12}$C$\rightarrow$3$\alpha$
 and $<P^*_{Ti}>\approx$200 MeV/  for $^{22}$Ne$\rightarrow$5$\alpha$
 (processing of the available data). There by we clearly see a tendency toward an increase
 of the mean $\alpha$ particle momentum with increasing their multiplicity. This implies a
 growth  of the total Coulomb interaction of alpha clusters arising in nuclei.\par
 \begin{figure}
    \includegraphics[width=5in]{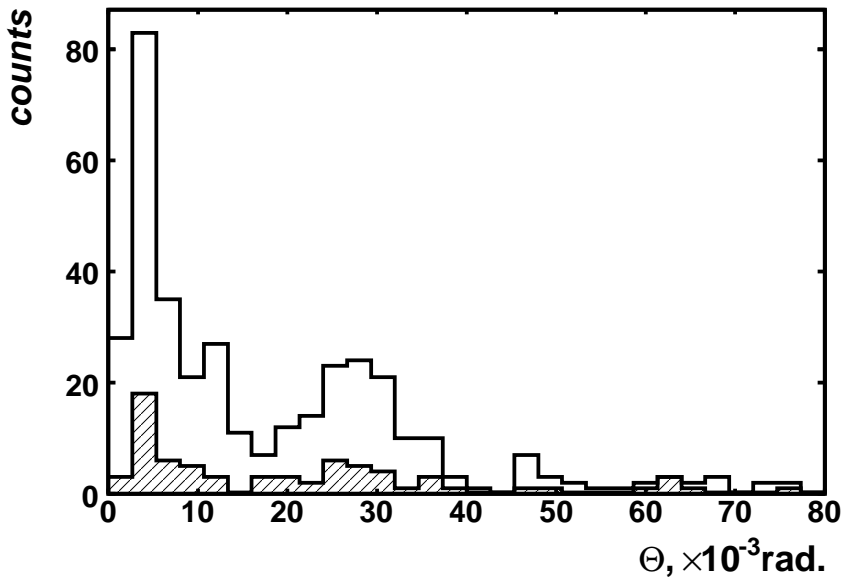}
    \caption{\label{fig:3} The opening $\Theta$ angle distribution of $\alpha$ particles
 in the $^9$Be$\rightarrow$2$\alpha$ fragmentation reaction at 1.2~A~GeV energy. The outer contour corresponds to all events.
 The inner histogram is obtained for events, which accompanied by protons recoil of emulsion
 target (dashed area).}
    \end{figure}
 \indent In the opening angle $\Theta$ distribution (Fig.~\ref{fig:3}) one can also see
 two peaks with mean values 4.6$\times 10^{-3}$rad. and 26.8$\times 10^{-3}$rad.
 The ratio of the numbers of the events in the peaks  is close to unity.\par
 \indent  The $\Theta$ distribution entails the invariant energy  Q$_{2\alpha}$
 distribution, which is calculated as a difference between the effective invariant mass
  M$_{2\alpha}$ of an $\alpha$ fragment pair and the doubled  $\alpha$ particle mass by
 the equations
\begin{eqnarray}
   {M^2_{2\alpha}=-(\sum_{j=1}^{2} P_j)^2} \nonumber\\
   {Q_{2\alpha}=M_{2\alpha} - 2\cdot m_{\alpha}}\label{eq3}
 \end{eqnarray}
where P$_j$ is the $\alpha$ particle 4-momentum.\par   
 \begin{figure}
    \includegraphics[width=5in]{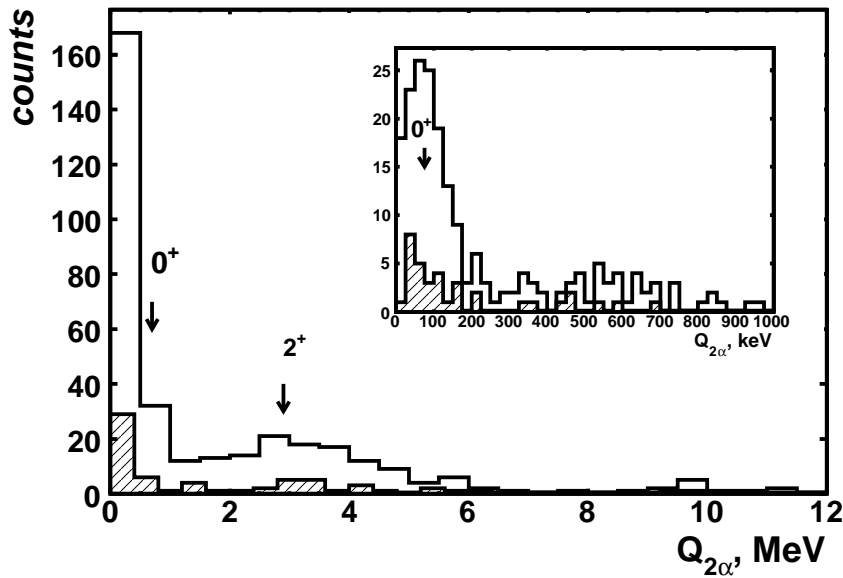}
    \caption{\label{fig:4}The invariant energy Q$_{2\alpha}$ distribution of $\alpha$
 particle pairs in the $^9$Be$\rightarrow$2$\alpha$ fragmentation reaction at 1.2~A~GeV energy.
  On the intersection: the Q$_{2\alpha}$ range  from 0 to 1~MeV.
 Arrows mark the $^8$Be nucleus levels (0$^+$ and 2$^+$). The outer contour corresponds
 to all events. The inner histogram is obtained for events, which accompanied by protons recoil of
 emulsion target (dashed area).}
 \end{figure} 
 \indent In the invariant energy Q$_{2\alpha}$ distribution (Fig.~\ref{fig:4}) there are two peaks in the ranges
  0 to 1 MeV and 2 to 4 MeV. The shape of the distribution does not contradict the suggestion
 about the $^9$Be fragmentation involving the production of an unstable $^8$Be nucleus which
 decays in the 0$^+$ and 2$^+$ states. The values of the peaks of the invariant energy
 Q$_{2\alpha}$ and the transverse momenta  P$^{*}_{T}$ in the c.m.s. relate to each other.
 To the Q$_{2\alpha}$ range from 0 to 1~MeV with a peak at 100~keV there corresponds a peak
  P$^*_{T}$ with  $<P^*_{Ti}>\approx$24~MeV/c , and to the Q$_{2\alpha}$ range from
 2 to 4~MeV  there corresponds a peak with $<P^*_{Ti}>\approx$101~MeV/c.\par
 
 \section{\label{sec:level3}Fragmentation of $^{14}$N nuclei}
 
\indent  A stack of layers of BR-2 emulsion was exposed to a beam of $^{14}$N nuclei
 accelerated \cite{arXiv14N} to a momentum of 2.86~A~GeV/c at the Nuclotron of the
 Laboratory of High Energy Physics (JINR). Already been found amoung 950 inelastic 
 events in which the total fragment charge was equal to the Z$_{0}$=7 fragment
 charge and there were no produced particles. Events were sought by viewing over the track
 length  which provided the accumulation of statistics without selection.
The selected events are divided in two classes. The events of the type of
 \lq\lq white\rq\rq star  and the interactions involving the production
 of one or a few target-nucleus fragments belong to the first class.\par
 \indent Table \ref{aba:tab1} shows the charge multi-fragmentation topology which was
 studied for the events satisfying the above-mentioned conditions. The upper line
 is the Z$>$2 fragment charge, the second line is  the number of single-charged
 fragments , the third one the number of two-charged fragments, and the fourth and
 fifth lines are the number of the detected events with a given topology for
 \lq\lq white\rq\rq stars and events with target-nucleus excitation for each
 channel, respectively. The two last lines present the total number of interactions
 calculated in absolute values and in percent.\par 
 \begin{table}
\caption{\label{aba:tab1}The charge topology distribution of the \lq\lq white\rq\rq stars and the
 interactions involving the target-nucleus fragment production in the $^{14}$N
 dissociation at 2.86~A~GeV/c momentum.}
\begin{tabular}{ l@{\hspace{5mm}}|@{\hspace{2mm}}c@{\hspace{2mm}}|@{\hspace{2mm}}c@{\hspace{2mm}}|
@{\hspace{2mm}}c@{\hspace{2mm}}|@{\hspace{2mm}}c@{\hspace{2mm}}|
@{\hspace{2mm}}c@{\hspace{2mm}}|
@{\hspace{2mm}}c@{\hspace{2mm}}|
@{\hspace{2mm}}c@{\hspace{2mm}}|@{\hspace{2mm}}c@{\hspace{2mm}}}
\hline\noalign{\smallskip}
\hline\noalign{\smallskip}
~Z$_{fr}$~& 6 & 5 & 5 & 4 & 3 & 3 & -- & -- \\
~N$_{Z=1}$~& 1 & -- & 2 & 1 & 4 & 2 & 3 & 1   \\
~N$_{Z=2}$~& -- & 1 & -- &  1 & -- & 1 & 2 & 3 \\
~N$_{W.S.}$~& 13 & 4 & 3 & 1 & 1 & 1 & 6 & 17  \\
~N$_{t.f.}$~& 15 & 1 & 3 & 3 & -- & 2 & 5 & 32  \\
~N$_{\sum}$~& 28 & 5 & 6 & 4 & 1 & 3 & 11 & 49   \\
~N$_{\sum, \%}$~& 26 & 5 & 5 & 4 & 1 & 3 & 10 & 46  \\
\hline\noalign{\smallskip}
\hline\noalign{\smallskip}
\end{tabular}
\end{table}
 \indent The analysis of the data of Table \ref{aba:tab1} shows that the number of
 channels involving Z$>$3 fragments for the \lq\lq white\rq\rq stars is larger by
 about a factor of 1.5 than that for the events accompanied by a target breakup.
 On the contrary, for the 2+2+2+1 charge configuration channel this number is
 smaller by about a factor of 1.5. Thus, in the events with target breakup, the
 projectile fragments more strongly  than in the \lq\lq white\rq\rq stars. The
 data of Table \ref{aba:tab1}  points to the predominance of the channel with the
 2+2+2+1 charge configuration (49 events) which has been studied in more detail. The obtained results show that the $^{14}$N
 nucleus constitutes a very effective source for the production of 3$\alpha$ system.\par
\indent In order to estimate the energy scale of production of 3$\alpha$ particle
 systems in the $^{14}$N$\rightarrow$3$\alpha$+X channel, we present the invariant
 excitation energy Q$_{3\alpha}$ distribution  with respect to the $^{12}$C ground state: 
\begin{eqnarray}
   {M^2_{3\alpha}=-(\sum_{j=1}^{3} P_j)^2} \nonumber\\
    Q_{3\alpha}=M_{3\alpha}^* - M(^{12}C)\label{eq4} 
  \end{eqnarray}
 where M($^{12}C$) is the mass of the ground state corresponding to the charge and the
 weight of the system being analyzed, M$^*_{3\alpha}$  the invariant mass of the system of
 fragments. Statistics was increased to 132 events $^{14}$N$\rightarrow$3$\alpha$+X
including 50 \lq\lq white \rq\rq stars by scanning over the emulsion plates.\par 
\begin{figure}
    \includegraphics[width=5in]{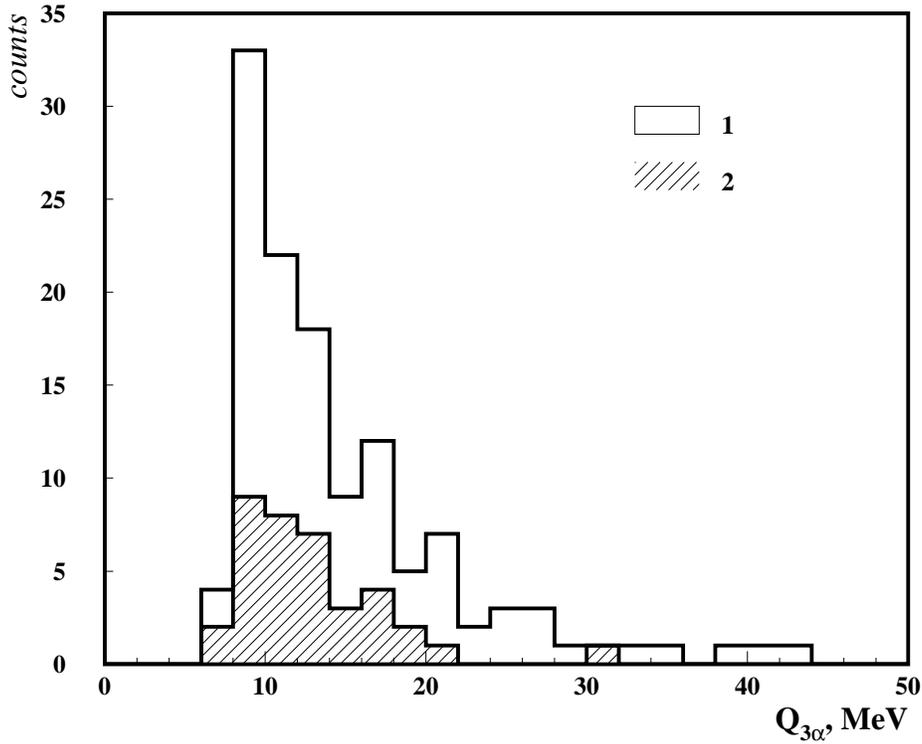}
    \caption{\label{fig:5}The invariant excitation energy Q$_{3\alpha}$
 distribution  of three $\alpha$ particles with respect to the $^{12}$C ground
 state for the process $^{14}$N$\rightarrow$3$\alpha$+X. The following notation
 is used$:$ 1) all the events of the given dissociation, 2) \lq\lq white\rq\rq stars.}
 \end{figure} 
 \indent The main part of the events is concentrated in the Q$_{3\alpha}$ area from 10 to
 14 MeV, covering the known $^{12}$C levels~(Fig. \ref{fig:5}). Softening of the
 conditions of the 3He + H selection, for which the target fragment production is
 allowed, does not result in a shift of the 3$\alpha$ excitation peak. This fact
 suggests the universality of the 3$\alpha$ state population mechanism.\par
 \begin{figure}
    \includegraphics[width=5in]{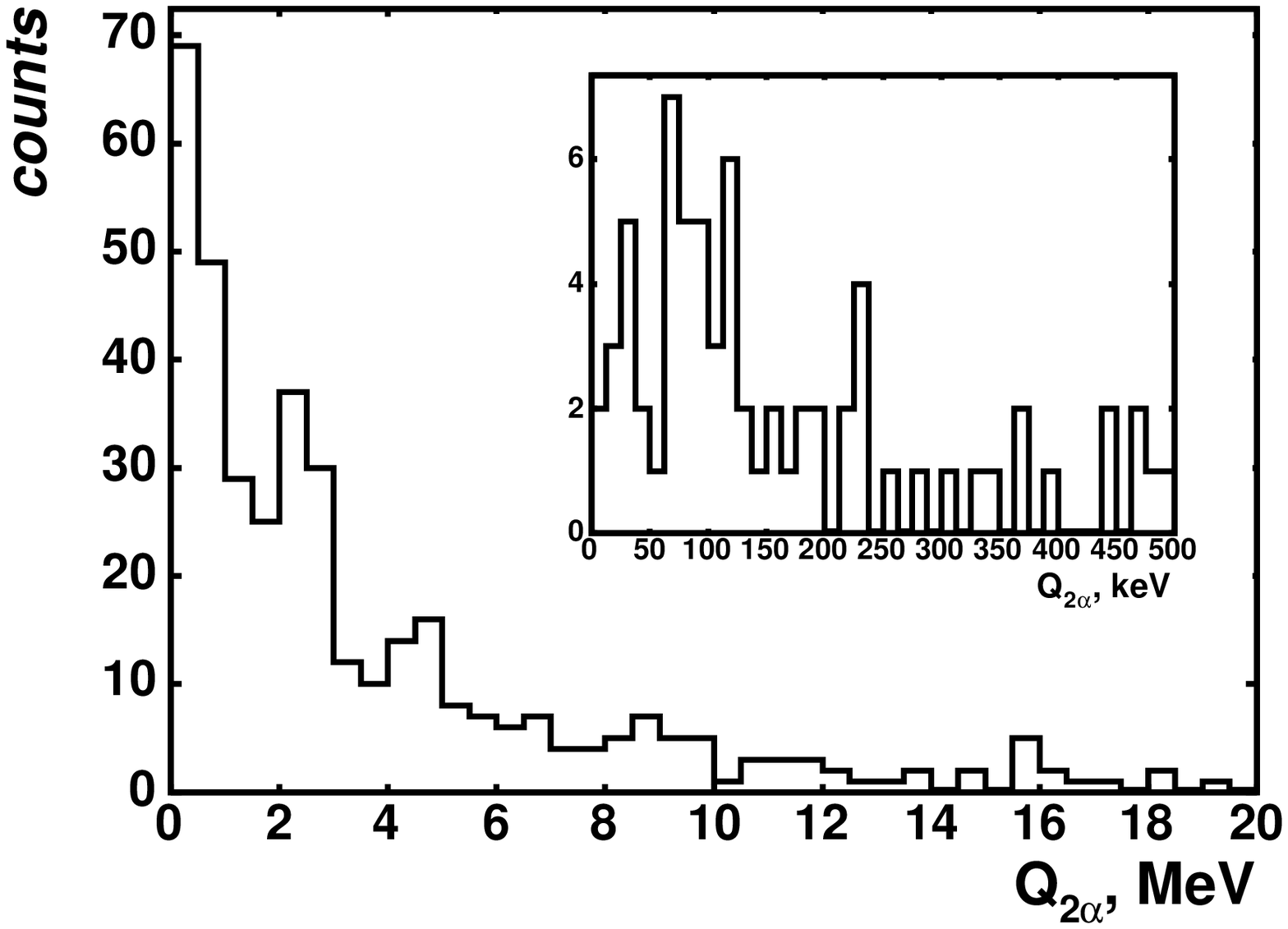}
    \caption{\label{fig:6} The invariant excitation energy Q$_{2\alpha}$ distribution
 of $\alpha$ particle pairs for the process $^{14}$N$\rightarrow$3$\alpha$+X. In
 the inset: a fraction of the distribution at 0-500~keV. }
    \end{figure}
\indent  To estimate the fraction of the events involving  the production of an
 intermediate $^8$Be nucleus in the reactions $^{14}$N$\rightarrow ^8$Be+X$\rightarrow$3$\alpha$+X
 we present the invariant excitation energy distribution for an $\alpha$ particle
 pair with respect to the $^8$Be ground state (Fig. \ref{fig:6}). The first
 distribution peak relates to the value to be expected for the decay products
  of an unstable $^8$Be nucleus in the ground state 0$^+$. 
The distribution centre is seen to coincide well with the decay energy of the
 $^8$Be ground state. The fraction of the $\alpha$ particles originating from the
 $^8$Be decay  is 25-30\%.\par
 
\section{\label{sec:level4}Fragmentation of $^7$Be, and $^8$B nuclei}

\indent The results of investigations dealing with the charge topology of the
 fragments produced in peripheral dissociation of relativistic $^8$B, $^7$Be nuclei in
 emulsion are presented in Ref~\cite{Web, arXiv1, arXiv2, arXiv3}.\par 
\begin{table}
\caption{\label{aba:tab2}$^7$Be fragmentation channel (number of events)}
\begin{tabular}{@{}c|c|c|c|c|c|c|c|c|c@{}}
\hline\noalign{\smallskip}
\hline\noalign{\smallskip}
Channel&2He&2He&He+2H&He+2H&4H&4H&Li+H&Li+H&Sum\\
~&n$_{b}=$0&n$_{b}>$0&n$_{b}=$0&n$_{b}>$0&n$_{b}=$0&n$_{b}>$0&n$_{b}=$0&n$_{b}>$0&~\\
\hline
~$^3$He+$^4$He&30&11&  &  &  &  &  &  & 41\\
~$^3$He+$^3$He&11& 7&  &  &  &  &  &  & 18\\
~$^4$He+2p    &  &  &13& 9&  &  &  &  & 22\\
$^4$He+d+p   &  &  &10& 5&  &  &  &  & 15\\
~$^3$He+2p    &  &  & 9& 9&  &  &  &  & 18\\
$^3$He+d+p   &  &  & 8&10&  &  &  &  & 18\\
~$^3$He+2d    &  &  & 1&  &  &  &  &  &  1\\
$^3$He+t+p   &  &  & 1&  &  &  &  &  &  1\\ 
~3p+d         &  &  &  &  & 2&  &  &  &  2\\
~2p+2d        &  &  &  &  & 1&  &  &  &  1\\
~$^6$Li+p     &  &  &  &  &  &  & 9& 3& 12\\
\hline
Sum          &41&18&42&33& 2& 1& 9& 3&149\\
\hline\noalign{\smallskip}
\hline\noalign{\smallskip}
\end{tabular}
\end{table}
 \indent Table \ref{aba:tab2} presents the numbers of the events detected in various channels of the $^7$Be fragmentation. Of them, the
$^3$He+$^4$He channel noticeably dominates, the channels $^4$He+d+p and $^6$Li+p constitute 10\% each. Two events involving
no emission of neutrons in the three-body channels $^3$He+t+p and $^3$He+d+d were registered. The reaction of 
charge-exchange of $^7$Be nuclei to $^7$Li nuclei was not detected among the events not accompanied by other secondary 
charged particles.The events involving no target fragments (n$_b$=0) are separated from the
events  involving one or a few  fragments (n$_b>$0).\par
\indent For the first time, nuclear emulsions were exposed to a beam of relativistic $^8$B nuclei. We have obtained data 
on the probabilities of  the $^8$B  fragmentation channels in peripheral interactions. 55 events of the peripheral 
$^8$B dissociation which do not involve the production of the target-nucleus fragments  and mesons 
(\lq\lq white\rq\rq ~stars ) were selected. A leading contribution of the $^8$B$\rightarrow^7$Be+p  mode having the lowest
 energy threshold was revealed on the basis of these events. Information about a relative probability of dissociation modes
 with larger multiplicity have been obtained. Among the found events there are 320 stars in which the total charge of the relativistic
fragments in a 8$^{\circ}$ fragmentation cone $\Sigma$Z$_{fr}$ satisfies the condition $\Sigma$Z$_{fr}>$3. These stars were attributed to the number 
of peripheral dissociation events N$_{pf}$.  The  N$_{pf}$ relativistic fragment distribution of over charges N$_Z$
is given in Table \ref{aba:tab3}. There are given the data  for 256 events containing the target-nucleus fragments  - N$_{tf}$,
as well as for 64 events which contain no target-nucleus fragments (\lq\lq white\rq\rq ~stars )-- N$_{pf}$. The role of the 
channels with multiple relativistic fragments  N$_{Z}>$2 is revealed to be dominant for the N\lq\lq white\rq\rq ~stars. 
Of peripheral events, the \lq\lq white\rq\rq ~stars N$_{ws}$ (Table \ref{aba:tab3}) are of very particular interest. They are not 
accompanied  by the target-nucleus fragment tracks and makes it possible to clarify the role of different cluster degrees
of freedom at a minimal excitation of the nuclear structure.\par
\begin{table}
\caption{\label{aba:tab3}The charge topology distribution of the number of  interactions of the peripheral 
N$_{pf}$ type (N$_{pf}$=N$_{tf}$+N$_{ws}$), which were detected in an emulsion exposed to a second $^8$B nucleus beam. Here Z$_{fr}$ is the total
charge of relativistic fragments in a 8$^{\circ}$ angular cone in an event, N$_{Z}$ the number of fragments with charge Z
 in an event, N$_{ws}$ the number of \lq\lq white\rq\rq stars, N$_{tf}$ the number of events involving the target fragments, 
N$_{ws}$ the number of \lq\lq white\rq\rq ~stars.}
\begin{tabular}{@{}c| c | c |c |c |c |c |c@{}}
\hline\noalign{\smallskip}
\hline\noalign{\smallskip}
~~Z$_{fr}$	~~&~~N$_{5}$~~&~~N$_{4}$~~&~~N$_{3}$~~&~~N$_{2}$~~&~~N$_{1}$~~&~~N$_{tf}$~~&~~N$_{ws}$\\
~~7	~~&~~-	~~&~~-	~~&~~-	~~&~~1	~~&~~5	~~&~~1  ~~&~~-\\
~~6	~~&~~-	~~&~~-	~~&~~-	~~&~~2	~~&~~2	~~&~~8	~~&~~2\\
~~6	~~&~~-	~~&~~-	~~&~~-	~~&~~1	~~&~~4	~~&~~6	~~&~~4\\
~~6	~~&~~-	~~&~~-	~~&~~-	~~&~~-	~~&~~6	~~&~~1	~~&~~-\\
~~5	~~&~~-	~~&~~-	~~&~~-	~~&~~1	~~&~~3	~~&~~61	~~&~~14\\
~~5	~~&~~-	~~&~~-	~~&~~-	~~&~~2	~~&~~1	~~&~~44	~~&~~12\\
~~5	~~&~~-	~~&~~-	~~&~~1	~~&~~-	~~&~~2	~~&~~8	~~&~~-\\
~~5	~~&~~-	~~&~~-	~~&~~1	~~&~~1	~~&~~-	~~&~~1	~~&~~-\\
~~5	~~&~~-	~~&~~1	~~&~~-	~~&~~-	~~&~~1	~~&~~17	~~&~~24\\
~~5	~~&~~1	~~&~~-	~~&~~-	~~&~~-	~~&~~-	~~&~~17	~~&~~1\\
~~5	~~&~~-	~~&~~-	~~&~~-	~~&~~-	~~&~~5	~~&~~21	~~&~~4\\
~~4	~~&~~-	~~&~~-	~~&~~-	~~&~~-	~~&~~4	~~&~~5	~~&~~1\\
~~4	~~&~~-	~~&~~-	~~&~~-	~~&~~2	~~&~~-	~~&~~24	~~&~~1\\
~~4	~~&~~-	~~&~~-	~~&~~-	~~&~~1	~~&~~2	~~&~~42	~~&~~-\\
\hline\noalign{\smallskip}
\hline\noalign{\smallskip}
\end{tabular}
\end{table} 
\begin{table}
\caption{\label{aba:tab4}The charged dissociation mode distribution of the \lq\lq white\rq\rq ~stars
 produced by the $^7$Be and $^8$B nuclei. To make the comparison more convenient, for the $^8$B nucleus
 one H nucleus is eliminated from the charged mode and  the  channel fractions are indicated.}
\begin{tabular}{@{}c |c| c| c| c@{}}
\hline\noalign{\smallskip}
\hline\noalign{\smallskip}
     ~~$\Sigma$Z$_{fr}$=4~~  & ~~$^7$Be~~& ~~\% ~~ & ~~$^8$B (+H)~~ & ~~\% ~~ \\
	~~2He~~	    & ~~41~~	& ~~43~~  & ~~12~~         & ~~40~~ \\
	~~He+2H	~~  & ~~42~~    & ~~45~~  & ~~14~~         & ~~47~~ \\
	~~4H~~	    & ~~2~~	    & ~~2~~	  & ~~4~~	       & ~~13~~ \\
\hline\noalign{\smallskip}
\hline\noalign{\smallskip}
\end{tabular}
\end{table}
 \indent Table \ref{aba:tab4} gives the  relativistic fragment charge distribution in 
the \lq\lq white\rq\rq ~stars for $^{7}$Be and $^{8}$B nuclei. The $^{8}$B events are presented without 
one single-charged relativistic fragment, that is a supposed proton halo. The identical fraction of the two main 2He and 
He+2H dissociation channels is observed for $^{7}$Be and $^{8}$B nuclei which points out that the $^{8}$Be core excitation 
is independent of the presence of an additional loosely bound proton in the $^{8}$B nucleus.\par

\section{\label{sec:level5} Conclusions}

\indent The degree of the dissociation of the relativistic nuclei in peripheral
 interactions can reach a total destruction into nucleons and singly and doubly
 charged fragments. The emulsion technique allows one to observe these
 systems to the smallest details and gives the possibility of studying them
 experimentally.\par
 \indent New experimental observations are reported from the emulsion exposures to
 $^{14}$N, $^9$Be, $^8$B, $^7$Be  nuclei with energy above 1~A~GeV. The main features of
 $^9$Be$\rightarrow$2He relativistic fragmentation are presented.
 For the particular case of the relativistic $^9$Be nucleus dissociation it is shown that precise
 angular measurements play a crucial role in the restoration of the excitation spectrum of
 the alpha particle fragments. This nucleus is dissociated practically totally through the 0$^+$
 and 2$^+$ states of the $^8$Be nucleus. The data obtained from $^9$Be angular measurements  can be employed for the
 estimation of the role of $^8$Be in more complicated N$\alpha$ systems.\par
\indent The results of the study of the dissociation of $^{14}$N nuclei of a primary
 momentum of 2.86~A~GeV/c in their interactions with the emulsion nuclei are also presented.
 The present investigation indicates the leading role of the 2+2+2+1 charge
 configuration channel. The energy scale of the 3$\alpha$ system production has been estimated.
  According to the available statistics 80\% of interactions are concentrated
 at 10-14~MeV. The fraction of the $^{14}$N$\rightarrow ^8$Be+X$\rightarrow$3$\alpha$+X
channel involving the production of an intermediate $^8$Be nucleus is about 25\%.\par 
\indent Advantages of emulsion technique are exploited most completely in the study
of peripheral fragmentation of light stable and neutron deficient nuclei.
 The results of investigations dealing with the charge topology of the fragments produced 
in peripheral dissociation of relativistic $^7$Be, $^8$B nuclei in emulsion are presented.
Information on the relative probability of dissociation modes with a larger multiplicity was
 obtained. The dissociation of a $^7$Be core in $^8$B indicates an analogy with that of the
 free $^7$Be nucleus.\par 

\end{document}